\documentclass[fleqn,twoside]{article}
\usepackage{espcrc2}

\usepackage{graphicx}
\usepackage[figuresright]{rotating}

\newcommand{\AmS}{{\protect\the\textfont2
  A\kern-.1667em\lower.5ex\hbox{M}\kern-.125emS}}

\title{The radiative return method - a short theory review}

\author{Henryk Czy\.z\address {Institute of Physics, University of Silesia, PL-40007 Katowice, Poland.}%
        \thanks{Work 
supported in part by EC 5th Framework Programme under contract
  HPRN-CT-2002-00311 
(EURIDICE network), TARI project RII3-CT-2004-506078 and
Polish State Committee for Scientific Research (KBN)
under contract 1 P03B 003 28}}
       
\begin{document}

\begin{abstract}
 A short review of the status of the theoretical developments 
 concerning the radiative return method is presented. 
 The emphasis is on the construction of the 
  PHOKHARA Monte Carlo event generator 
  and its tests. It is advocated that the radiative return method
 provides not only with the hadronic cross section extraction
 competitive with and complementary to the scan method, but also
 that it is a powerful tool in detailed studies of the hadron interactions.
\vspace{1pc}
\end{abstract}

% typeset front matter (including abstract)
\maketitle
\newcommand{\bea}{\begin{eqnarray}} 
\newcommand{\eea}{\end{eqnarray}}
\newcommand{\non}{\nonumber}
\section{INTRODUCTION}
The hadronic cross section measurement is crucial for the
accurate evaluation of the hadronic contributions to the 
 muon anomalous magnetic moment ($a_\mu$) \cite{Passera} and running
 of the electromagnetic coupling $\alpha_{QED}$ \cite{Fred}.
The traditional way of measuring of the hadronic cross section
 via the energy scan has one disadvantage - it needs dedicated
 experiments.  An alternative way, the radiative return method,
 was proposed in \cite{Zerwas},
 even if the radiative process was investigated earlier \cite{Fadin}.
 This method, described in the next section, allows for a simultaneous
 extraction of the hadronic cross section from the nominal energy
 of the experiment down to the production threshold, and importantly can
 profit from the data of all high luminosity meson factories.
\section{THE HADRONIC CROSS SECTION VIA THE RADIATIVE RETURN METHOD}
  The radiative return method
  relies on an observation that the cross section
 of the reaction $e^+e^- \to \mathrm{hadrons} +\mathrm{photons}$,
 with photons emitted from the initial leptons,
 factorizes into a function $H$,
  fully calculable within QED, and 
 the cross section of the reaction $e^+e^-\to \mathrm{hadrons}$
\bea 
&&\kern-30ptd\sigma(e^+e^- \to \mathrm{hadrons} + \gamma\mathrm{'s})(s,Q^2) =
 \non \\  \kern+30pt &&H \cdot
 d\sigma(e^+e^-\to \mathrm{hadrons})(Q^2) \ ,
\label{master}
\eea
where $Q^2$ is the invariant mass of the hadronic system.
 Thus from the measured differential, in $Q^2$, cross
 section of the reaction $e^+e^- \to \mathrm{hadrons} + \mathrm{photons}$
 one can evaluate $\sigma(e^+e^-\to \mathrm{hadrons})$ once the function $H$
 is known.
As evident from the Eq.(\ref{master}), the radiative return method allows
for the extraction of the hadronic cross section from the 
 production energy threshold of a given hadronic channel
 almost to the nominal energy of a given experiment ($\sqrt{s}$).  
 The smaller cross section of the radiative process as compared
 to the process without photons emission has to be compensated
 by higher luminosities. That requirement is met by meson factories
 (DAPHNE, BaBar, BELLE). All of them were built for other
 purposes then the hadronic cross section measurements, but their
 huge luminosities provide with data samples large enough
 for very accurate measurements of interesting hadronic
 channels and/or give an information on rare channels, which were
 not accessible in scan experiments. Two representative examples
 of such measurements are the very accurate pion form factor
 extraction by KLOE collaboration \cite{KLOE2} and 
 $\sigma(e^+e^-\to 3\pi)$ extraction by BaBar collaboration \cite{babar3pi},
 where it was shown that the old DM2 scan data were wrong at high
 values of $Q^2$. An extensive review of the recent results
 of both collaborations concerning the radiative return is presented in 
 \cite{Achim}. If one likes to use the formula Eq.(\ref{master}) in a realistic
 experimental situation, where sophisticated event selections are
 used, one needs a Monte Carlo event generator of the measured process.
 To meet that requirement the PHOKHARA \cite{Rodrigo:2001kf} event generator
 was constructed.
\section{THE PHOKHARA MONTE CARLO EVENT GENERATOR AND ITS TESTS}
 The construction of the PHOKHARA event generator started from the
 EVA generator \cite{Binner:1999bt,Czyz:2000wh},
 where structure function method
 was used to model multi-photon emission.
  The physical accuracy of the program was however
 far from the demanding experimental accuracy of the KLOE pion form factor
 measurement and in a series of papers that high expectations were met.
 The first distributed PHOKHARA version \cite{Rodrigo:2001kf} relied
 on the one loop initial state radiative corrections calculated 
 in \cite{Rodrigo:2001jr} and the two hard photon emission was simulated
 using exact matrix element written within helicity amplitudes method.
 That version was designed to run with tagged photon configurations
 and the radiative corrections necessary for a photon emitted at small
 angles were calculated afterwords in \cite{Kuhn:2002xg} and implemented
 into the event generator in \cite{Czyz:2002np}. The important issue
 of the final state emission, which will be discussed in details in 
 the next section, was addressed in \cite{Czyz:PH03} and subsequently in
 \cite{Czyz:PH04}, while in aspects specific for $\phi$-factory DAPHNE
 in \cite{Czyz:2004nq}. In parallel the generator was being extended to allow
 for the generation of more hadronic channels and now it allows for
 generation of 
 $\pi^+\pi^-$, $K^+K^-$, $\bar K^0K^0$, $\bar p p$, $\bar n n$,
  $\pi^+\pi^-\pi^0$, $2\pi^+2\pi^-$, $\pi^+\pi^-2\pi^0$ hadronic states
  and $\mu^+\mu^-$.
  The nucleons final states were discussed in \cite{Nowak}, while
 the three pion current was modeled and implemented into the PHOKHARA
 event generator in \cite{Czyz:2005as}.

 All that allowed for building of the state-of-the-art
 event generator. The proper implementation of the radiative corrections
 as well as the hadronic currents is guarantied by extensive tests
 of the generator discussed below.

  At each step of
 the generator development, the newly implemented matrix element
 calculated in the program using the helicity amplitude method,
 squared and summed over all available helicities
 is compared with the square of the matrix element summed over 
 polarizations calculated by traditional trace method.
 All numerical calculations in PHOKHARA are performed using
 double precision, but in some cases,
 mainly for double photon emission it was necessary to use the quadrupole
 precision for the matrix element evaluation calculated analytically
 by the trace method.
 It was caused by numerical 
 cancellations up to ten significant digits occurring between various terms.

  Another type of tests concern the process of the generation.
 The initial state emission of one photon with one-loop radiative corrections
 and two hard real photon emission from initial states were compared
 \cite{Czyz:2002np} separately
 with existing analytical results for fully inclusive phase space
 configurations \cite{Berends:1988ab}. Both results were in perfect
 agreement up to the numerical precision of the tests limited
  by the Monte Carlo statistics. The relative difference between
  the numerical and analytical results  was a few times
 $10^{-4}$, that was well within the statistical
 Monte Carlo error bars. That accuracy is usually called
 technical precision of the Monte Carlo generator and I will use
 that name hereafter. That tests are repeated for each newly
 added hadronic channel, even if the program uses the same
 building blocks for every channel, to avoid possible bugs
 in the implementation. 
 
 Similar tests were performed for
 the reaction $e^+e^-\to\pi^+\pi^-\gamma(\gamma)$ with one photon
 emitted from the initial state and one
 emitted from the final pions \cite{Czyz:PH03}. In that case
 the analytical results of \cite{Schwinger:ix} were used together
 with the analytical results obtained in \cite{Czyz:PH03} to test
 the Monte Carlo generation. Again the same technical precision was achieved
  and the tests were repeated for the charged kaons in the final states.
 For the reaction $e^+e^-\to\mu^+\mu^-\gamma(\gamma)$ the configurations
 with one photon emitted from the initial states and
  one from final muons were tested in \cite{Czyz:PH04}
  with the same technical precision.
 
 It would be useful to make extensive comparisons with independent
 Monte Carlo generators, however the only 
 existing  Monte Carlo code meeting the accuracy requirements is the 
 KKMC \cite{Jadach:1999vf}, which is limited to
 muons in the final state as far as its accurate matrix element is concerned.
 It means that one can tests the initial state emission and in fact
 detailed tests were performed \cite{Jadach:KKMC} leading to an
 excellent agreement of the non exponentiated matrix elements
 of the virtual corrections to single photon emission 
 (a relative difference of a few times $10^{-5}$
 was found). The higher order effects, that can be seen as a difference
 between the exponentiated and the 
 non exponentiated matrix elements reach at most
 2 per mile with the exception of the region of the invariant mass
 of the hadronic system very close to the nominal energy of the experiment.
 That region, where soft multi-photon emission play an important role
 and thus the exponentiation is necessary,
  is however of no interest to the radiative return method. All that results
 agree very
 well with the estimated previously in \cite{Rodrigo:2001kf},
 by means of the structure 
 function approach,
   PHOKHARA physical precision of 0.5\%, attributed to the
 lack of the higher order effects in the ISR matrix element.
\section{THE FINAL STATE EMISSION}
 The final state emission (FSR) forms a potential problem for the application
 of the radiative return method and it has to be studied carefully
 to be sure that the required accuracy of the description is met.
 First of all one has to have in mind that the situation at 
 B- factories is completely different from the one of the $\phi$- factory
 DAPHNE. 
 In the former case the  region of
 hadronic masses, which is of physical interests, mainly
 below 4~GeV, lays far from the nominal energy of the
 experiments, thus an emission of a hard photon is required to
  reach it. As a result the typical kinematic configuration of an 
 event consists of a photon emitted in one direction and hadrons
 going opposite to it. That provides a natural suppression of
 the FSR contributions, which are large for photons emitted parallel
 to the direction of a charged hadron in the final state, and makes
 the measurement of the hadronic cross section easier. For the $\phi$- factory,
 where the  physically interesting region is not far from the nominal
 energy of the experiment, that natural separation between the emitted
  photon and the hadrons does not exists and one has to suppress FSR
 by an appropriate event selection. In that case one has to control
 the uncertainty due to the model dependence of the final state emission.
 That is a challenge, as the models were not tested with the adequate
 precision prior to the DAPHNE results. I will discuss that problem
 on the basis of the $e^+e^-\to \pi^+\pi^-\gamma(\gamma)$ process,
 where the accuracy requirements are the most demanding. For that process
 the solution of the problem was first proposed in \cite{Binner:1999bt}
 and further elaborated in \cite{Czyz:PH03}. One can imagine a similar
 solution for other hadronic final states, but that kind of analysis
 was never performed. 
 
 The main tool in the tests of the model(s) of the photon
 emission from the final pions is the charge asymmetry.
 For one real photon emission the two-pion state is produced in C=-1 and
 with an 
 odd orbital angular momentum for the real photon emitted from the initial
 state and in C=1 and with an even orbital angular momentum for the 
 real photon emitted from the final state. As a result, the initial-final
 state interference is odd under $\pi^+\leftrightarrow \pi^-$ interchange
 and it integrates to zero for charge blind event selections. In the same time
 it is the only source of the charge asymmetry and as such allow for
 tests of the models of the final state emission. The charge asymmetry
  depends on the invariant mass of the two-pion system and that
 allow for deeper insight into details of the tested model(s). 
 In short, the tests should be done in the following way: First
 one compares the experimental data  for the asymmetry with the Monte Carlo
 where the tested model was implemented. That has to be performed 
 for an event selection which enhance the FSR as compared to the ISR.
 Once the implemented model agrees with the data one chooses an 
 event selection, which suppresses the FSR and performs the 
 radiative cross section measurement. That guaranties that the
 ISR and the FSR contributions are separately well under control.
 For the case of untagged photons a specific background,
 $e^+e^-\to\pi^+\pi^-e^+e^-$, has to be also taken into account
 \cite{Hoefer:2001mx,Czyz:2006dm} as the final leptons are not vetoed.
 
 The  reaction
 $e^+e^-\to \pi^+\pi^-\gamma$,
  with the photon emitted from the pions, does contribute also to 
 dispersion integrals for evaluation of $a_\mu$ and $\alpha_{QED}$
 and in the former case its theoretically estimated value \cite{Czyz:PH03}
 is of the size of the theoretical uncertainty and thus numerically
 important. As its theoretical estimations are not reliable it has 
 to be measured.
 The sketched program was successfully undertaken
 by KLOE and resulted in a sound extraction of the 
 $\sigma(e^+e^-\to \pi^+\pi^-)$ \cite{KLOE2} together
 with the mentioned FSR photon corrections.
 
  Another source of complications for using of
 the radiative return method
 at DAPHNE are the radiative $\phi$ decays. That problem
 was considered for the first
 time in \cite{Melnikov:2000gs} and it is discussed in more details 
 in the next section.
\section{THE RADIATIVE RETURN AS A TOOL IN HADRONIC PHYSICS}
 The reaction $e^+e^-\to\phi\to\pi^+\pi^-\gamma$  produces the same final
 state as the one measured for the  $\sigma(e^+e^-\to \pi^+\pi^-)$
 extraction. That contribution is sizable for energy close to the
 $\phi$ mass and thus important for DAPHNE. As shown in \cite{Czyz:2004nq},
 the charge asymmetry has large analyzing power and can provide with
 information allowing for distinguishing between different models
 of the radiative $\phi\to\pi\pi\gamma$ decay, even if that is
 impossible in the  analysis of the differential (in $Q^2$) cross section.
  Again  by an appropriate
 event selection one can suppress those contributions or enhance them
 as for other sources of the FSR emission discussed in the previous section.
  That example shows that the radiative return method can be used
 not only for the hadronic cross section measurement, but also
 for getting detailed information about the models of hadronic
 interactions. It was partly exploited in the KLOE analysis 
\cite{Ambrosino:2005wk},
 however as the charge asymmetries were not used in the fits, the
 collected data contain more information on the tested models then
 actually was used.
 
An extensive analysis of the FSR contributions is extremely important 
 especially in the ongoing  KLOE analysis, both for tagged \cite{Debora}
 and untagged \cite{Federico} photon(s) as till now the experimental
 information on the pion-photon interactions is far from being
 satisfactory and it is not clear if the model used currently in PHOKHARA
 (sQED + vector dominance + radiative $\phi$ decays) describes 
 the FSR with the adequate precision in the threshold region,
 where other contributions might be important \cite{Pancheri:2006cp}.
  
 Another example of the power of the radiative return method
 in the hadronic models tests 
 is the separation of the magnetic and the electric nucleon form factors.
 The method, which was proposed in \cite{Nowak}, was used 
 by BaBar collaboration \cite{Aubert:2005cb} for separation of the proton
 form factors.
 The obtained results show clearly its competitiveness.
\section{THE SUMMARY AND NEAR FUTURE DEVELOPMENTS}
\begin{figure}[htb]
\includegraphics[height=6cm,width=0.9\linewidth]{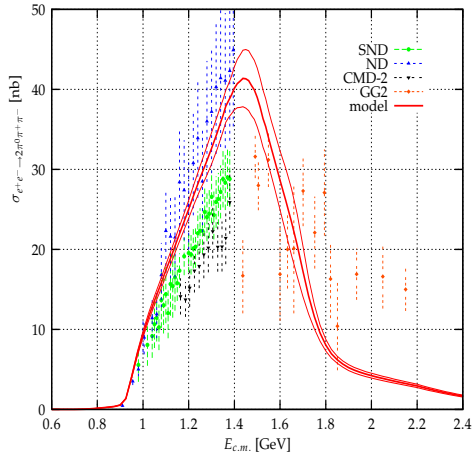}
\caption{The predictions for the $\sigma(e^+e^-\to2\pi^0\pi^+\pi^-)$
 (central dashed line). The lower and upper dashed lines
 show the error bars of the model predictions. }
\label{fig:4pi}
\end{figure}

 A short description of the theoretical status of the radiative return
 method was presented showing its competitiveness in precise measurements
 of the hadronic cross section and
 studies 
 of the hadronic interactions. Many interesting problems, for example
 a proper modeling of the hadronic current of 
 multi-meson final states observed at BaBar \cite{Achim},
  the FSR simulation for more than the two-pions final states,
 the modeling of the narrow resonance contributions
  and many others not mentioned
 in this paper still await for detailed theoretical investigations.
One of that problems, which will be addressed 
in the near future by the group working on the PHOKHARA event generator
  developments and updates, is the improvement of the theoretical
 description of the $4\pi$ hadronic current \cite{Wapienik}. 
 Exploiting isospin symmetry and all available experimental data one comes
 to the predictions for the $\sigma(e^+e^-\to2\pi^0\pi^+\pi^-)$
 (central dashed line)
 shown in Fig.\ref{fig:4pi}. The lower and upper dashed lines
 show the error bars of the model predictions.

{\bf Acknowledgments:}
 The author thanks the organizers 
of the The International Workshop  "e+e- collisions from $\phi$ to $\psi$" for 
hospitality and stimulating atmosphere of the whole event. 
\vspace{-0.5cm}

%%%%%%%%%%%%%%%%%%%%%%%%%%%%%%%%%%%%%%%%%%%%%%%%%%%%%%%%%%%%%%%

\end{document}